\documentclass{article}

\usepackage{arxiv}

\usepackage[utf8]{inputenc} 
\usepackage[T1]{fontenc}    
\usepackage{hyperref}       
\usepackage{url}            
\usepackage{booktabs}       
\usepackage{amsfonts}       
\usepackage{nicefrac}       
\usepackage{microtype}      
\usepackage{lipsum}
\usepackage{graphicx}
\usepackage{floatrow}

\title{Deep Learning Based Android Malware Detection Framework}

\author{
 Soumya Sourav \\
  Department of Computer Science\\
  University of Texas at Dallas\\
  Dallas, TX, USA 75080\\
  \texttt{sxs180011@utdallas.edu} \\
  \And
  Devashish Khulbe \\
  Center for Urban Science + Progress\\
  New York University\\
  New York, USA 10003\\
  \texttt{dk3596@nyu.edu} \\
  \And
  Naman Kapoor \\
  Department of Computer Science\\
  Delhi Technological University\\
  New Delhi, India 110042 \\
  \texttt{naman.kapoor12@gmail.com} \\
}

\begin{document}
\maketitle

\begin{abstract}
With the development in the field of smartphones and ever growing base of Internet, various softwares are left prone to many malicious activities like pharming, phishing, ransomware, spam, spoofing, spyware, eavesdropping, etc. These threats have not spared the smartphones which are equally prone to them. In this work, we aim to detect these malwares with accuracy and efficiency. This being essentially a classification problem, we use various machine learning methods for this task. We observe that across models, Attention based Artificial Neural Networks (ANN), or broadly speaking, Deep Learning, are most suitable for this problem. Attention based ANNs are an amalgamation of accuracy and efficiency, the crux of our work. The accuracy achieved by our model is around 96.75\%. Our model runs the test on Android Package Files (APKs) to determine whether a particular application is malicious or not by doing behavior analysis on android application under consideration.
\end{abstract}


\section{Introduction}
[1] The recent ubiquitousness of mobile phones for doing each and every task in day to day life because of there increased functionality has left them vulnerable to many new malware threats. Along with the nature of mobile devices, which are upgrading each day, the mobile threats are also upgrading with each passing day. The term mobile now includes all kinds of devices of IoT. Threats can be in the form of banking trojans, ransomwares, spywares, etc. which often leads to stealing private data and money from the user. Malware detection has continued to be one of the biggest challenges in today’s technological world considering the fact that the resources available for doing the same are very limited. [17] According to McAfee malware analysis report 2017 the unprecedented rise of ransomwares is expected to continue in 2017. Google has removed more than 4000 apps from playstore in the past year without any information to the users. According to the Telemetry data obtained by McAfee Mobile Threat Research, more than 500,000 mobile users still have these apps installed in their devices thus making them vulnerable to the threats these apps pose . [15] Kaspersky Mobile security products detected 1,319,418 malicious installation packages, 28,796 mobile banking trojans, 200,054 mobile ransomwares. TrojanBanker has become one of the most easiest and most used malware by fraudsters.AndroidOS.Asacub is one of the malwares most prominently used. The Q2 2017 results of the Kaspersky Lab were equally threatening with malware detection being at 1,319,148 which is twice the previous quarters data. In Q2 2017, Adware with a contribution of 13.31\%, was the biggest source of malwares. The share increased by 5.99\%. The majority of all discovered Installation packages are detected as AdWare.AndroidOS.Ewind.iz and AdWare. AndroidOS. Agent.n. Trojan-SMS malware (6.83\%) ranked second in terms of the growth rate: its contribution increased by 2.15 percentage points. 
   The problem of using machine learning based classifier to detect malwares using android  permissions presents 3 main challenges: first, we need to extract an apps main features which will provide the base to classify them; second, important features should be identified and taken into account from the obtained dataset.; third, classification using the model.
    The first problem can be solved using various reverse engineering tools like androguard, apk tool, etc. which extracts an APK file to give its permissions, API calls and other related information about the file. We have used androguard in this project. The second problem is solved using Principal Component Analyses for Logistic Regression, Gradient Boosting and Naive Bayes and [4] sklearn.ensemble.ExtraTreesClassifier (ETC) for Neural Network. ETC is a meta estimator that fits a number of randomized decision trees (a.k.a. extra-trees) on various sub-samples of the dataset and use averaging to improve the predictive accuracy and control over-fitting. The third problem is solved using separate models out of which the Artificial Neural Network model that consists of an eleven-layered structure with 105 number of nodes in each layer gives the highest accuracy.
   This paper is organization augments understanding at every level: in Section II, we have discussed the related works in this field; in Section III, we have explained the working of our model; in Section IV, we have presented the components of our methodology; in Section V, we have discussed our experimental results; \& finally, in Section VI, we have provided the references of our project .

\section{Related Work}
The machine learning methods for classification [1,2,3,4,5,8] has been very prevalent among data scientists and researchers. With some changes in their framework some encouraging results have been obtained. 
   Another work [2] used the ensemble features to categorize among the malware apps. They have utilized the androguard as well for breaking down the android permissions along with mining the API calls and software/hardware features. The project is implemented in two phases: first one includes using separate features and the other one using the combined or the ensemble features. The accuracy was around 93\% for the ensemble technique.
   One other work [4,9] has used android permissions for their projects as well. However, k-nearest classifier [4] is used in one and random forest tree classifier [9] is used in another. Both are very robust and powerful machine learning algorithms.
   By analyzing the number of times each system call has been issued during execution of certain task, Crowdroid [6], a machine learning-based framework, recognizes Trojan-like malware. The trojanized version of an application is very different from its genuine application, since every time it is used, it issues different types of and different number of API calls. 
   The project [7] makes use of parallel functioning of machine algorithms for storing different features. The algorithms used are Rule Based Classifier, Function Based Classifier, Tree Based Classifier, Probabilistic Classifier. All the features from these algorithms, after they are trained, are combined and a net result is obtained whether an app is malicious or not. The proposed method generates some good results, however the inclusion of four different algorithms surely speeds it down.



\section{Methodology}

\subsection{Part 1: Data \& Features}
First phase of the work involves reading dataset, which includes permissions and API calls of various apps which are classified whether they are malwares or benign applications. Relevant features are then selected using two separate techniques: Principal Component Analyses (PCA) and a wrapper classifier by segregating features w.r.t. their importance. Irrelevant features are then removed for better generalization and better accuracy as well as reduced time involved in classification.

\subsubsection{Dataset}

   The dataset used for analysis is provided by Kaggle, a competitive platform where data miners and scientists can obtain numerous datasets and can compete in various challenges hosted by this domain only. Generally companies and researchers post some data on it and data miners compete to provide the best predicting accuracy on that data. The dataset contains features of 338 applications which are labelled as ‘benign’ or ‘malicious’. This label is used to do feature selection and classification analysis by supervised learning. The dataset comprises of a .csv file. The top row contains all the permissions. The presence of a specific permission in an application sample is indicated by 1 and 0 if it is not present.
   
  Imbalance problem - the situation when the contents of each class are not proportionate- is a very important aspect when using binary classifiers for the detection of malicious code. In our case the absence of malwares and legitimate files in equal proportions causes the imbalance problem. It is need of hour to continuously update the training set with new malicious files for making the classifier better. This is really an important aspect in maintaining such a framework. The dataset is updated to accommodate for new samples using androguard, an opensource project to extract features from APKs. This solves for the imbalance problem by maintaining the proportion of benign and malicious entries to be equal.
  
\subsubsection{Features}
A feature is a significant estimable property or characteristic of an event under observation. Each attribute in the dataset set signifies a unique feature and each entity signifies a data sample (an android application in this case). A feature in this project corresponds to a permission specified by an application. The dataset contains 330 features. The feature vector contains feature data in binary form where ‘1’ indicates the presence of a permission and ‘0’ indicates absence.
   The malware analysis approach implemented in this paper uses static android features extracted from the android app which are used to determine if app contains any malicious permissions which depends on a trained classification model. Androguard does the job of extracting features to train the ANN model combination of malware and benign APKs.
   Three categories of features are used for learning phase: 1. App Permissions 2. API calls 3. Standard OS and framework commands.
   API calls are the calls which are made to the server in the name of an application using a SDK or an API. Android permissions are requests or permissions acquired by the apps in order to use certain system data and features to maintain security for the system and user.
   
\begin{table}
 \caption{Overview of the features extracted from the apps.}
  \centering
  \begin{tabular}{lll}
    \toprule
    \cmidrule(r){1-2}
    Type     & Features \\
    \midrule
    API call related & abortBroadcast, getDeviceId, \\ 
                      & getSubscriberId, getCallState, getSimSerialNumber, \\
                      & android.provider.Contacts, android.provider.ContactsContract; \\
                    & HttpUriRequest, SMSReceiver, bindService, \\ & onActivityResult, LjavaxCryptoSpecSecret, DexClassLoader, 
                    \\
                    & getNetworkOperator, getSimOperator \\
    Command related     & .apk; pmsetInstallLocation; pminstall; \\
    & GET-METADATA; GET-RECEIVERS; \\ 
    & GET-SERVICES; GET-SIGNATURES; \\
    & GET-PERMISSIONS \\
    Permissions     & android.permission.UPDATE-LOCK;\\
    & android.permission.USER-ACTIVITY;\\
    & android.permission.VIBRATE;
android.permission.WAKE-LOCK; \\
    & android.permission.WRITE-CALENDAR;\\
    & android.permission.WRITE-CALL-LOG;\\
    & android.permission.WRITE-CONTACTS;\\
    & android.permission.ACCESS-FINE-LOCATION

  \end{tabular}
  \label{tab:table}
\end{table}

\subsubsection{Feature Extraction}
Overfitting is an issue where the model instead of describing the underlying relationship, describes noise and random error. An overfitted model tends to perform very poorly on predictive tasks as it overreacts to minor fluctuations in training data. This is a very common problem with every machine learning model. Huge and complex datasets affect training models by reducing the efficiency of the classification problem by using unnecessary number of features resulting in increased computation time. Hence extraction of important features is a key part before modeling.
We use two methods for selection of important features:

\paragraph{Principal Component Analysis}
Principal Component Analysis (PCA) [12] is an important technique to deal with multicollinearity in the data and eliminating redundant variables. It drops the “least important” variables while still retaining the most valuable parts of all of the variables. As an added benefit, each of the new variables after PCA are all independent of one another. 
In our data set, the original feature space of 330 variables is reduced to 23 variables after using PCA. This transformed feature space accounts for 90\% of the variance in the original space.

\paragraph{ExtraTreeClassifier}
ExtraTreeClassifier is an amalgamation of a search technique, which selects the features, and an evaluation measure, used to score the feature subsets. The algorithm varies in complexity with some being as simple as testing all the possible subsets of selected features and then winnowing the one with minimal error rate. One of the most influential factor for the algorithm is the choice of evaluation metric and these evaluation metrics are the one’s which create a distinction among three main feature selection algorithm: filters, embedded methods and wrappers.
This technique makes use of wrapper method to figure pout the accuracy of the selected features’ subsets. By counting the number of mistakes our model makes while predicting on the hold-out set, the accuracy can easily be determined. Wrapper methods tend to be very intensive when it comes to computations. However the best results on a model can be obtained by this method only.

\subsection{Part 2: Modeling}
Second part involves training the machine learning models using the processed dataset and saving weights and the trained model.
We use Logistic Regression, Naive Bayes and Gradient Boosted Trees along with the artificial neural network and compare their performance on the test data. We use the truncated feature space after the PCA and ExtraTreeClassifier segregates the useful variables. The test dataset is dynamic and keeps on changing whenever the program is run. It consists of around 35-40 permissions belonging to the original dataset with almost 330 permissions and generated on 398 applications both benign and malicious. This models after being trained, saves the weights and the model. These weights and model are then called which breaks individual apps to be checked and then the neural network model decides whether the app is malicious or benign.\\
The crux of our work is developing the Artificial Neural Network for this problem and compare it with other off the shelf techniques. Thus, we focus on the results of ANN for major parts of the modeling and results section.

\subsubsection{Classification Methods}
We begin with off the shelf models for the classification task. We use Logistic Regression, Naive Bayes, Random Forests and Gradient Boosting models before moving onto deep neural networks.
These methods are fairly well known and have proven to perform good for binary classification problems. The problem with these approaches, however, is that they might not be able to learn complex non-linearity which might be in the data. To tackle this, we delve into developing Artificial Neural Network for the task.

For the Artificial Neural Network we add an attention layer to enhance the prediction capability of the neural network.
The neural architecture has 12 Dense layers.
In between every third Dense layer a dropout layer was added to reduce the overfitting of the neural networks. The dropout layer drops a certain percentage of the neurons in random. These neurons don't receive any updates and don't contribute to the output at all.
The attention network \cite{b10,b11} is used in order to make the neural network capable of attending to a subset of the inputs which increases the space of functions which can be approximated by the neural network and makes it possible to look into entirely new use-cases which enhances the neural net's capability to converge to a global minimum. The attention network generally assigns the scores to the input features using the softmax function.
\begin{equation}
f(z)\textsubscript{j} = \frac{e^{z_{i}}}{\sum _{j=1}^{K}e^{z_{i}}}
\end{equation}

where f\textsubscript{j} is the output of attention layer and e\textsubscript{j} is the output previous Dense layer. This output is the probability distribution which is combined with the Dense layer output which gives the importance of each feature for the output prediction. Different methods for combining the output layer and probability distribution can be used like Dot product, Scalar Multiplication or Bilinear combination. For the final training procedure of the PKG and PCB model, a batch size of 32 was fixed after doing cross validation with the optimizer "Adam" was used.

\section{Results}
For analyzing the performance of our models, we decided to check their accuracies on the test data. Checking the accuracy makes sense as the distribution of positive and negative records in our data are equal in proportion. The training:test data proportion is set to be 75:25. 

On comparing the accuracy across all models, ANN, due to its complex structure, is able to perform better than all of the other classification techniques. The validation accuracy is found to be 95\%. The problem of classification could be countered with many machine learning algorithms like SVM or regression. However, the main advantage Neural Networks or the multilayer perceptrons hold over these algorithms are the fixed number of layer size. The neural networks are parametric models while the machine learning algorithms are not i.e. in neural networks, there are a number of hidden layers depending upon the number of features. The outputs of neural networks can be multiple and they were tested to be faster than the machine learning algorithms.
\begin{table}[h]
 \caption{Classification accuracy across models}
  \centering
  \begin{tabular}{lll}
    Name     & Accuracy(\%) \\
    \midrule
    Naive Bayes     & 88.3 \\
    Random Forests & 90.8 \\
    Gradient Boosting     &  91 \\
    Logistic Regression & 93.6 \\
    Attention Based ANN & 95.3\\
    \bottomrule
  \end{tabular}
  \label{tab:table}
\end{table}

We also check the performance of our ANN model with respect to its predictions concerning the individual classes through a confusion matrix.

\begin{figure}[h]
\centering
\begin{minipage}{.5\textwidth}
  \centering
  \caption{Performance of ANN model: Confusion matrix (left) and accuracy curve (right)}
  \includegraphics[width=.8\linewidth]{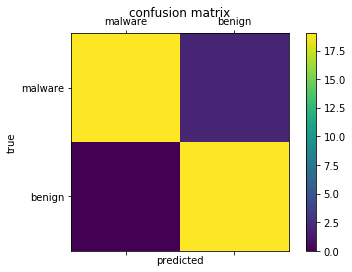}
  
  \label{fig:test1}
    \end{minipage}%
    \begin{minipage}{.5\textwidth}
  \centering
  \includegraphics[width=.8\linewidth]{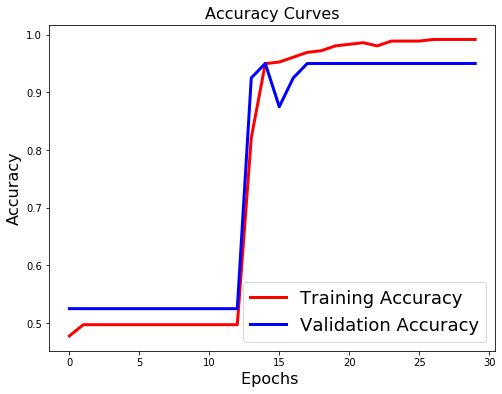}
  \label{fig:test2}
\end{minipage}
\end{figure}

Precision is a common metric to judge the performance of classification models. It is essentially the fraction of relevant examples classified correctly. With the ANN model, the precision score is found to be 0.90476. 

\begin{equation}
    ANN Precision = \frac{True Positive}{True Positive + False Positive} = 0.90476
\end{equation}

\section{Conclusion}
We introduced a new Deep Learning based method to predict malwares in Android applications which performs significantly better than traditional off the shelf machine learning techniques. Due to the small size of the data we worked on, this new method is yet to be tested on bigger and more complex data with more features and across various platforms. This can be a possible scope for future work. Also, we fed only the significant features from our original data into the models' inputs. There can be further scope in finding which features are important predictors for this task. Furthermore, we think malware detection is a fairly important area of study and different approaches and data sets still need to be explored in this domain.

\bibliographystyle{unsrt}  


\end{document}